\documentclass[usenatbib]{mnras}

\usepackage[T1]{fontenc}
\usepackage{ae,aecompl}

\usepackage{graphicx}	
\usepackage{amsmath}	
\usepackage{amssymb}	

\newcommand{\nc}{\newcommand}
\nc{\teff}{$T_{\rm eff}$\,}
\nc{\logg}{\rm log\,$g$\,}
\nc{\kms}{\,${\rm km\,s}^{-1}$\,}
\nc{\mic}{$\xi_{\rm t}$\,}
\nc{\ms}{~m~s$^{-1}$\,}
\nc{\cms}{~cm~s$^{-1}$\,}
\nc{\sms}{sub-m~s$^{-1}$\,}

\title[Data Reduction Pipeline of the TOU Optical Very High Resolution Spectrograph]{Data Reduction Pipeline of the TOU Optical Very High Resolution Spectrograph and Its \sms Performance}

\author[Ma \& Ge (2018)]{Bo Ma$^{1,2}$ 
\thanks{E-mail:mabo8@mail.sysu.edu.cn} and
 Jian Ge$^{1}$ \thanks{E-mail:jge@astro.ufl.edu} \\
$^{1}$Department of Astronomy, University of Florida, 211 Bryant Space Science Center, Gainesville, FL, 32611-2055, USA \\
$^{2}$School of Physics and Astronomy, Sun Yat-sen University, Zhuhai 519082, People's Republic of China \\
}
\date{Accepted XXX. Received YYY; in original form ZZZ}

\pubyear{2018}

\begin{document}
\label{firstpage}
\pagerange{\pageref{firstpage}--\pageref{lastpage}}
\maketitle

\begin{abstract}
TOU is an extremely high resolution optical spectrograph (R=$100,000$, 380-900~nm), which is designed to detect low 
mass exoplanets using the radial velocity technique.
We describe an IDL-based radial velocity (RV) data reduction pipeline for the TOU spectrograph 
and its performance with stable stars. This pipeline uses a least-squares fitting algorithm to 
match observed stellar spectra to a high signal-to-noise template created for each star. 
By carefully controlling all of the error contributions to RV measurements in both the hardware and 
data pipeline, we have achieved $\sim$0.9\ms long-term RV precision with one of the most RV stable stars, 
Tau Ceti, similar to what has been achieved with HARPS. This paper presents steps and details 
in our data pipeline on how to reach the \sms RV precision and also all major error 
sources which contribute to the final RV measurement uncertainties. 
The lessons learned in this pipeline development can be applied to other environmentally 
controlled, very high resolution optical spectrographs to improve RV precision.
\end{abstract}


\section{INTRODUCTION} 
Discovery of an abundant population of exoplanets is one of the most exciting developments in 
astronomy in the last two decades, and radial velocity (RV) technique has made significant contributions to 
these discoveries. 
To date, RV surveys have detected $\sim$500 exoplanets \citep[RV planets from Exoplanet.org;][]{han14}, which show 
an fascinating diversity in terms of masses, orbital periods, and eccentricities distributions, 
from the short period hot Jupiters, to planets in very elongated orbits, to planetary systems with multiple Jupiter-mass planets, 
to massive planets in very close binaries, and to close-in super-Earths with periods of less than 100 days 
\citep{butler04, mcarthur04, santos04, rivera05, lovis06, udry06, howard10, mayor11, howard14, ma16}. 
New observations from innovative new RV instruments will continually reveal the unanticipated diversity of planetary systems.

Motivated by the investigation of the close-in super-Earth populations around nearby FGKM dwarfs and 
the search for habitable super-Earths, we developed a compact, 
extremely high resolution optical spectrograph with a broad wavelength coverage, called TOU from 2010$-$2013, 
and commissioned it at the 2 meter Automatic Spectroscopic Telescope \citep[AST,][]{eaton04a, eaton04b}, 
a robotic telescope at Fairborn Observatory in Arizona in 2013 July. TOU is a fiber-fed, cross-dispersed echelle 
spectrograph with a spectral resolution of about 100K, a wavelength coverage of 3800$-$9000~\AA, 
and a 4k$\times$4k Fairchild CCD detector \citep{ge12, ge14a, ge16}. 
This instrument holds a very high vacuum of less than 1 micro torr and about 1mK (rms)
temperature stability over a few months. We carried out a pilot 
RV survey of 20 nearby bright FGK dwarfs ($V<9$) at AST between 2014$-$2015
with the goal of fine tuning performance of the TOU spectrograph and achieving \sms RV precision.
Thanks to the unique design of a vacuum-sealed chamber, high mechanical stability, 
and accurate thermal control, the instrument is very stable. The measured wavelength solution on each pixel of 
the CCD shows nearly no sign of changing during each observing night. 
This spectrograph is designed to potentially reach better than 0.3\ms long-term RV precision. 

Given that a few next generation RV stabilized spectrographs targeting 0.1\ms are under construction or being 
commissioned \citep[such as ESPRESSO/ESO;][https://www.eso.org/public/usa/news/eso1739]{pepe13} and the significant 
investment in hardware development to achieve higher RV stability, it is critical that the data 
analysis tools are also as optimal as possible. There are three popular algorithms aiming at computing 
RVs, including a direct cross-correlation with a numerical binary mask \citep[hereafter the CCF technique;][]{baranne96, pepe02}, 
an optimized least-square fit to a spectral template \citep[hereafter the template-matching technique;][]{anglada12, zechmeister17}, 
and the forward modeling spectral synthesis technique \citep[hereafter the forward-modeling technique;][]{butler96}. The first two techniques 
are generally chosen for RV instruments with stabilized point spread functions such as  HARPS 
\citep[High-Accuracy Radial Velocity Planetary Searcher;][]{mayor03}. According to \citet{pepe02} 
and \citet{anglada12}, the template-matching technique is more precise at extracting RV information 
from stellar spectra than the CCF technique, especially for M dwarfs. 
However, the CCF technique is faster for RV data extraction than the template-matching one  
after an observations is concluded because there is no requirement to acquire 10$-$20 more 
stellar spectra from the same star to combine and generate a high signal-to-noise ratio (SNR) 
spectral template for RV extraction as required in the latter. The forward-modeling technique is 
generally applied to RV instruments that have their wavelength calibration sources located within 
the light paths of target stars, which can be used to model the line spread function variations of the instrument 
and effectively track and correct instrument drifts. For example, Keck/HIRES uses an 
iodine gas cell as its wavelength calibration source \citep{butler96}, and use it to model its 
instrument line spread function variations using extreme high resolution ($\rm >1000K$) spectra 
from the iodine gas cell to obtain high precision RV measurements after instrument RV drifts are corrected.

We use the template-matching technique for the TOU RV data pipeline.
This paper presents all the steps in our data processing and RV extraction pipeline, which 
has achieved \sms precision. 
It also includes lessons learned during the pipeline development. 
This knowledge can help pipeline development for the next generation of stabilized RV instruments 
and help reveal main areas that need to be improved in order to reach 0.1\ms RV precision.
The paper is organized as follows. We first introduce our RV observations and the basic algorithms 
used in the TOU pipeline in Section 2. Section 3 reports our instrument on-sky 
performance results for one stable star. In Section 4 we present the detailed error budget analysis for each step of
the TOU data reduction pipeline. In Section 5 we present the summary of the paper and discuss possible paths 
toward achieving 0.1\ms Doppler RV precision.

\section{Data Reduction and Radial Velocity Measurement }
\subsection{Observations}
\begin{figure}
\includegraphics[angle=0,width=8.2cm]{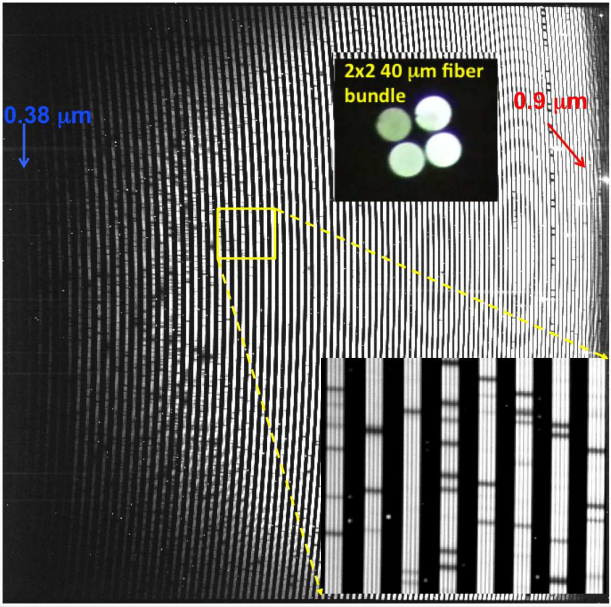}
\caption{The CCD image of Tau Ceti  with a 10~min exposure time using the TOU spectrograph. 
The spectrum is taken by the 2-m Automatic Spectroscopic Telescope at Fairborn observatory in 2014 November. 
The horizontal (X-axis) and vertical axes (Y-axis) are corresponding to the spatial and dispersion directions, respectively. 
} 
\label{fig:ccd}
\end{figure}

We obtained stellar spectra with the TOU spectrograph at
the 2-m AST at Fairborn observatory between 
2014 and 2015. The exposure time for a typical 
star ranges from 10 to 30~min to obtain data with sufficient SNRs 
($>100$ per pixel at 5500~\AA) and to average out short-term high frequency stellar oscillations 
\citep{mayor03, dumusque11}. Under good weather conditions, we can reach a 
SNR$\sim$100 per pixel for a V=6 mag G-type dwarf at 5500~\AA~with a 15minute exposure.
The CCD image of a bright G-type RV stable star, Tau Ceti, with a 10~min exposure time using TOU 
is displayed in Figure~\ref{fig:ccd}. The spectral orders are closely aligned in 
the vertical direction of the CCD while different spectral orders are separated horizontally 
by a two-prism cross-disperser \citep{ge12, zhao12}. We use a 2$\times$2 40$\micron$ fiber 
bundle to couple with the 80 micron diameter fiber from the telescope and slice the input 
image four times and rearrange the four sliced images in a linear row at the spectrograph 
entrance to double the spectrograph spectral resolving power \citep{ge12}. This fiber 
image slicer is the key to enable the compact design of the spectrograph while achieving 
the required very high spectra resolution of $\rm R=100,000$. We took daily master 
calibration frames in the afternoon. The master calibration data obtained every night 
include bias frames, flat-fielding frames using a tungsten lamp, and wavelength 
calibration frames using two Thorium-Argon (ThAr) lamps. One ThAr lamp is 
called the `reference' lamp, while the other is called the `operation' lamp. Following 
the HARPS operation, we use calibration data from both ThAr lamps to calibrate 
RV measurements. The reference lamp was turned on once per day to take 
calibration frames for wavelength calibration and instrument long-term drift 
tracking while the operation lamp was on for the entire night for tracking 
intra-night instrument drifts. Because the TOU spectrograph is extremely 
stable with typical instrument drifts less than 1\ms each night, we only took operation 
ThAr data at the beginning, the middle, and the end of each night to track 
intra-night instrument drifts.

\subsection{Spectrum Extraction}
\begin{figure}
\includegraphics[angle=0,width=8.2cm]{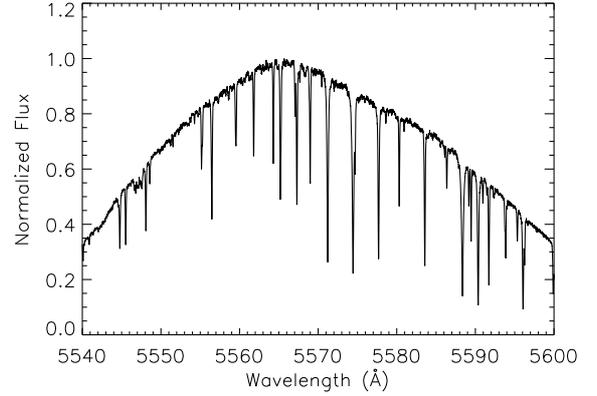}
\caption{One dimensional optimally extracted TOU spectrum of Tau Ceti, before removing the blaze function of the order. 
In our RV extraction pipeline, we do not normalize the spectrum using its continuum as this process is 
not necessary in our RV extraction as described in section 2.5.} 
\label{fig:1d}
\end{figure}

In the next several sections, we will describe the pipeline steps taken to transfer stellar exposure CCD images 
to RV data. The detailed error budget will be presented in section~\ref{sec:error}.
We reduce the TOU CCD spectral images using a standard optimal 
extraction technique \citep{horne86, piskunov02, zechmeister14} to trace spectral orders 
on the two dimensional echelle CCD image, rectify the orders, and then sum pixel flux in columns to 
obtain a one-dimensional spectrum for each order. The standard bias subtraction, 
scattered-light subtraction, and flat-fielding are also included in this extraction step. 
We use a master bias image created from a combined frame from ten 0~s exposures 
taken before each night's observation for the bias subtraction.
Scattered-light are modeled by fitting a three order polynomial function to the 
inter-order background flux among the 10 nearest orders on both sides of the order we want to extract, 
then interpolate to obtain the background flux values on the location of the current working order.
Flats with different exposure times are used to optimize the SNR in the blue and red 
part of the CCD, because the tungsten lamp has far more flux at longer wavelengths than at shorter ones. 
The flat fielding spectrum is constructed from a large number of Tungsten lamp
spectra (each with a median SNR $>$ 400 per pixel) extracted in the same way as stellar spectra obtained in the same day. 
The spectral profile in the X-direction (the spatial direction) is modeled using high SNR 
Tungsten lamp exposures on each night for summing the two dimensional (2D) stellar spectra
and create the final extracted one dimensional (1D) stellar spectra. Cosmic rays are also rejected 
at the same time when one dimensional spectra are extracted. 
Figure~\ref{fig:1d} shows part of the reduced 1D spectrum of Tau Ceti, which has not been de-blazed.



\subsection{Wavelength Solution}

We created a wavelength solution by first developing a code to automatically identify stable 
Thorium lines in our Thorium-Argon (ThAr) calibration spectra and fit these lines with a Gaussian function to 
derive the line central pixel values. We use a line list similar to that from \citet{redman14} since the ThAr lamps 
used in our TOU spectrograph are from the same manufacturer as those used in \citet[][private communication]{redman14}\footnote{ThAr lamps were manufactured by Photron Pty Ltd (Part no. P858A), which are metallic-thorium lamps. } 
We will upload our line list online for others to use. 
This has been verified after we took high resolution spectra of two of our ThAr lamps using the 2-m Fourier 
Transform Spectrometer (FTS) at the National Institute of Standards and Technology (NIST) in August 2016. 
We then make a high precision two-dimensional wavelength solution by fitting 
the line centroid pixels, spectral orders, and wavelengths of these stable Thorium lines. 
The wavelength solution model used is as follows: 
\begin{equation}
\lambda_{o,i} = a_{o,0}+a_{o,1}*i+a_{o,2}*i^{2}+a_{o,3}*i^{3},
\end{equation}
where $o$ is the order number and $i$ is the pixel number, $a_{o,0}$, $a_{o,1}$, $a_{o,2}$, and $a_{o,3}$ are 
the wavelength solution coefficients for order $o$. We fit this wavelength solution model 
using the identified Thorium lines. We were able to identify about $2000$ Thorium lines 
in the 4000$-$6200~{\AA} wavelength range. The wavelength residuals for all the Thorium lines with 
respect to the fitting values have an rms of $0.0008$~\AA. 

To test the repeatability of the wavelength solution, we did an experiment. In this experiment, we first 
created five wavelength solutions from ThAr exposures obtained in the same hour. Then we compute the RVs 
of one bright stellar exposure using these wavelength solutions. The RVs have a scatter of $\sim$0.3\ms, 
which demonstrates the error of the `zero-point' of the wavelength solution. We want to note that the 
relative stableness of this `zero-point' is important when masking telluric lines, but not as critical when 
deriving differential RVs. 

\subsection{Building a Stellar Template}

We build a stellar template with the highest possible SNR for each star in order to deliver an optimal RV 
precision. In the first pass of all the spectra, the preliminary RV for each epoch 
is obtained with respect to a preliminary stellar template using the same algorithm described in section~\ref{sec:match}. 
The preliminary stellar template is chosen as the stellar spectrum with the highest SNR. 
These preliminary RV measurements are then used to shift stellar spectra at each epoch and combine them 
to create a final stellar template. Specifically, spectra obtained in different epochs are not sampled at the same wavelength grid and 
cannot be co-added directly without interpolation. The preliminary template is served as a 
reference to generate a grid of reference wavelength in each order. 
Each stellar spectrum is then interpolated to the reference wavelength grid using a cubic spline function. 
The spline interpolation used in our pipeline is not flux conserving. However, this interpolation has an 
negligible effect on RV measurements based on our simulations.
The final stellar template flux value on each re-sampled wavelength grid is computed using a 3 
sigma clipped weighted mean over all epochs. Telluric absorption features deeper than $1\%$ of the continuum level 
are masked out during this co-adding process since these features do not co-move together 
with stellar absorption lines and would 
compromise the quality of the template.


\begin{figure}
\includegraphics[angle=0,width=8.2cm]{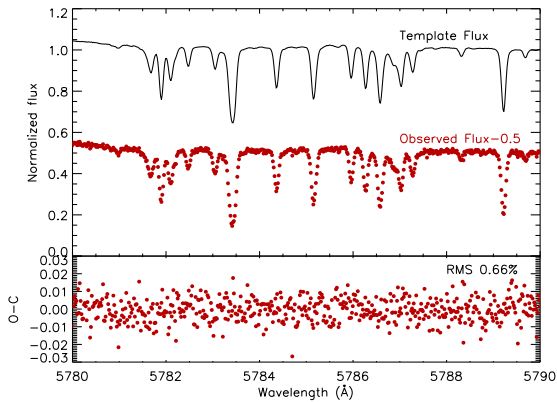}
\caption{ A spectrum matching result from a 10~{\AA} chunk of Tau Ceti spectrum. The spectra 
shown here are normalized for display purposes. The SNR per pixel of 
the observation is 173, which corresponds to a residual rms of 0.58\% assuming that 
the observed spectrum only contains photon noises and the spectral template has a negligible 
amount of photon noises. Since the template is generated by combining over one hundred Tau Ceti spectra, 
it has a very high SNR per pixel ($>1000$) and negligible amount of photon noises. The actual spectrum matching 
residuals shown in the bottom panel have an rms value of 0.66\%. The slight excess of residual 
rms comparing with photon noises suggests that assuming Poisson statistics is usually slightly optimistic.
\label{fig:match}} 
\end{figure}

\subsection{Global Continuum and Template-Matching}
\label{sec:match}
The RVs are derived using a method that we call the global continuum and template-matching 
technique, which is similar to that used in \citet{anglada12} and \citet{zechmeister17}. 
In this process, we use the RV shift of the star, the blaze function derived from the Tungsten 
lamp calibration spectra, and a three order polynomial function to match an observed 
spectrum to the template using a non-linear Levenberg-Marquardt least squares technique.
The initial guesses of the RVs are estimated from a 400 pixel trunk in the order 110 using 
a fast cross-correlation algorithm \citep{anglada12}, which are typically within 100\ms of the 
final RVs, to minimize the required number of computationally expensive least-squares iterations. 

The shape of the continuum from the Tungsten lamp spectrum in each spectral order 
is largely determined by the combination of the Tungsten intrinsic spectrum and the 
echelle grating blaze function. Due to the temperature difference between a star 
observed and the Tungsten lamp, their black body radiation spectrum continuum shapes 
are different. The continuum shape difference between two black bodies in a narrow 
wavelength range covered by each spectral order can be approximately described as 
a low-order polynomial function \citep{brahm17}. To measure relative RVs, we only 
need to know the variation of this low-order polynomial from one stellar exposure 
to the next exposure, which can be modeled using another low-order 
polynomial \citep{anglada12}. This use of the new low-order polynomial correction 
can also account for instrumental/observational effects, such as the impact of 
airmass and atmospheric dispersion/extinction, and wavelength-dependent fiber 
coupling \citep{anglada12, zechmeister17}. \citet{anglada12} investigated the 
impact of the choices of this new polynomial function on the RV measurements 
using HARPS data and found that a three order polynomial is optimal in their 
pipeline \citep[see also][]{zechmeister17}. In our data pipeline, we also use 
a three order polynomial to correct wavelength-dependent flux variations 
between different stellar exposures, which are uncorrected by the Tungsten calibration spectra. 



The parameters that yield the best fit between the template and the observed spectrum are derived using the
standard Levenberg-Marquardt least-squares algorithm. 
The adopted uncertainty in the determined RV is derived from the square root of the variance of the best fit.
Figure~\ref{fig:match} illustrates this matching result 
for a small trunk of the spectrum of Tau Ceti. The rms of the residuals is a little higher than the rms estimated from the 
Poisson photon noise distribution, which suggests there are other errors involved in this matching process, e.g., an imperfect template. 

\subsection{Barycentric Velocity Correction}

To calculate the differential RV with respect to the barycentric reference, 
we also need to correct the observer's heliocentric motion against the barycenter of the solar system, 
the Earth's rotation, and the influence of all the other bodies in the solar systems. 
The barycentric velocity is derived for each stellar exposure using the 
flux weighted centroid exposure time and code from \citet{wright14}. 
The flux weighted centroid time is calculated from the Photomultiplier Tube (PMT) counts recorded 
simultaneously during a stellar exposure using $\sim$4$\%$ of the photons from the science fiber output. 

\subsection{Instrumental RV Drift Correction}
%

\begin{figure}
\includegraphics[angle=0,width=8.2cm]{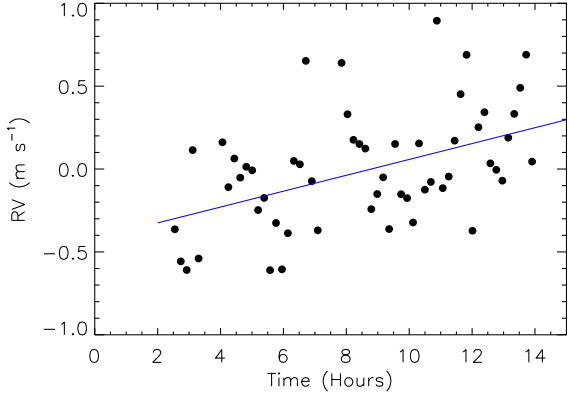}
\caption{Intra-night instrument drifts on a typical night tracked with the ThAr `operation lamp'.  
We use a linear function to fit these measurements from 
the `operation lamp'. The rms of the fitting residuals is $\sim$0.3\ms.
\label{fig:drift2}
} 
\end{figure}

The final correction to apply in our RV pipeline is the instrumental RV drift correction 
using reference ThAr frames taken in the afternoon and operation ThAr calibration 
frames taken every night in-between stellar exposures. 
Figure~\ref{fig:drift2} shows typical intra-night instrument 
drifts, which are less than 1~\ms. Since the TOU instrument is very stable in 
each night, we can interpolate these measured instrumental RV drifts at the beginning, middle and end of the night 
to derive the drift correction for each stellar exposure. 
The drifts measured by reference frames are used for correcting the instrument long-term drifts in stellar RV measurements.  


\section{On-Sky RV Performance}

\begin{figure}
\includegraphics[angle=0,width=8.2cm]{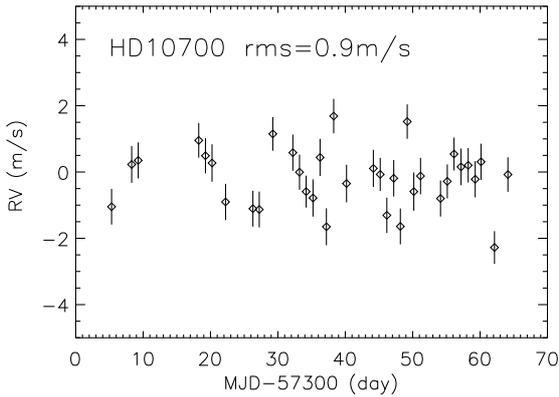}
\caption{Radial velocity measurements of an RV stable star, Tau Ceti, using the TOU spectrograph.
} 
\label{fig:tau_ceti}
\end{figure}

The short-term velocity rms is a good metric for characterizing instrumental precision.
Tau Ceti (HD~10700, $V=3.5$, G8V) is known as one of the most 
RV-stable bright G dwarfs \citep{pepe11}, which makes it an ideal target for the purpose of 
calibrating our processing pipeline. Thus, we observed this target near meridian three times 
each night within a 40~min window whenever the weather was good, for a total of 35 nights over 60 days. 
Most of the exposures were acquired at an airmass between 1.4$-$1.7.
Each exposure is 10 minutes to ensure that high SNR (SNR$\sim$330 per resolution element, including 3.3~pixels, at 5500~\AA~on average) 
spectra are collected. Figure~\ref{fig:tau_ceti} shows these RV measurements, which have 
an rms value of $\rm{0.9~m/s}$. Each RV measurement is derived by combining three consecutive 
10 min exposures on each night, which can effectively minimize the high frequency stellar oscillation noise. 
We only applied instrument drift correction using Th-Ar exposures for these RV measurements.
This rms is consistent with that from HARPS measurements reported by \citep{pepe11}. 
The RV rms before we combine the three data points on each night is $\sim$1.3\ms. Sources which contribute 
to this RV measurement error are further analyzed and reported in the following section.


\section{RV Error Budget Analysis}
\label{sec:error}
 
In this section we discuss all error sources that are relevant to the \sms RV precision 
from the observations, instrument, and data reduction pipeline. All error sources discussed 
in this section are summarized in Table~\ref{tbl:budget}, which have a combined value of 0.92\ms. 
This total estimated error is consistent with our on-sky measurement performance.

\begin{table}
  \caption{Error budget for Tau Ceti with SNR$\sim$300 per pixel at 5500~\AA.}
  \begin{tabular}{ c  c}
    \hline
    Error Sources &  RV Error \\ 
        & (m~s$^{-1}$) \\ 
    \hline
Photon Noise  &    0.27 \\
Wavelength Calibration  &    0.4 \\
Micro-Telluric Lines Contamination &    0.1$-$0.2 \\
Instrument Drift Correction  &    0.36 \\
Barycentric Velocity Correction  &   $<0.05$ \\
Template Interpolation  &  $<0.1$   \\
PSF Temporal Variation  &  0.2 \\
Flat-Fielding Variation & $<0.1$ \\
Charge Transfer Inefficiency & $<0.1$ \\
Color Effect & 0.06 \\
Stellar Jitter  &     0.45 \\
Telescope Guiding & $0.3$    \\
    \hline
Total Combined Error  &  $\sim$0.88 \\ 
    \hline
    \label{tbl:budget}
  \end{tabular}
\end{table}

\subsection{Photon Limited Errors}
We compute here the errors in the RV measurements due entirely to the poisson noise of the 
spectral images \citep{butler96, bouchy01}. We used two methods to calculate the error values: one using a
theoretical method by calculating the `Q' factor \citep{bouchy01} and the other using a simulation method by 
putting simulated spectra with poisson noise through the pipeline.

For the theoretical `Q' factor method, the photon-limited RV error is given by: 
\begin{equation}
\sigma_{\rm photon} = \frac{c}{Q\sqrt{N} },
\end{equation}
where N is the total number of photons over the entire spectral range, c is the speed of 
light, and Q is the quality factor of the stellar spectra.

\begin{figure}
\includegraphics[angle=0,width=8.2cm]{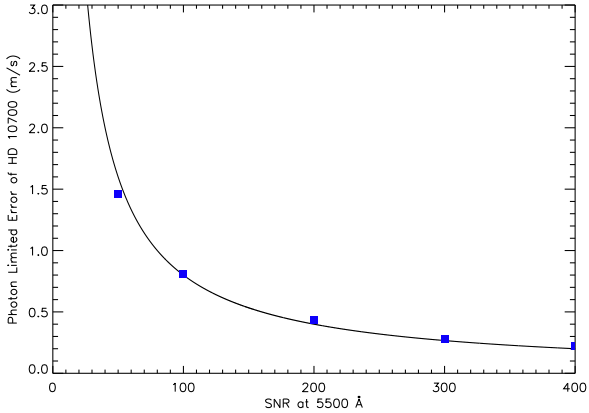}
\caption{Photon limited RV errors for Tau Ceti using the TOU spectrograph at 4250$-$6200~\AA. The 
blue diamonds are results from our simulation and the solid  curve is calculated 
from the `Q' factor. } 
\label{fig:photon_noise}
\end{figure}

In the simulation method, we generate simulated spectra of Tau Ceti by adding poisson noises to the 
stellar template. All the simulated Tau Ceti spectra have an input RV of 0\ms. 
The wavelength coverage of the stellar template is $\sim$4250$-$6200~\AA. 
We then calculate RVs for these simulated spectra and the standard deviation of the final 
RV's distribution is treated as the photon-limited error, $\sigma_{\rm photon}$.  
Figure~\ref{fig:photon_noise} displays the $\sigma_{\rm photon}$ results for Tau Ceti 
as a function of SNR from both methods, which shows a linear inverse relation as expected. 
With a SNR of 300 per pixel at 5500~\AA, the photon limited RV error is $\sim$0.33\ms for 
Tau Ceti, which is included in Table~\ref{tbl:budget}.

\subsection{Wavelength Calibration Errors}
\begin{figure}
\includegraphics[angle=0,width=8.2cm]{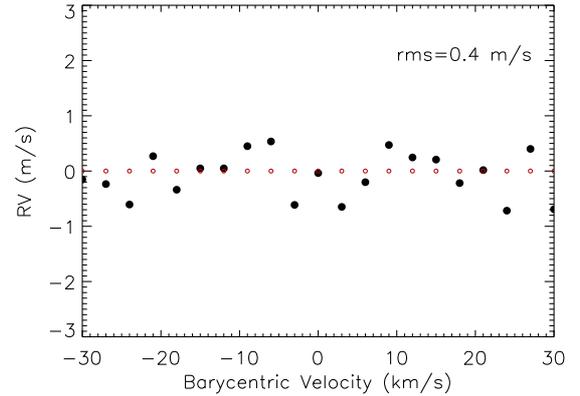}
\caption{  Simulation study of RV errors induced by the wavelength solution errors. 
In the simulation, we calculate the RV by matching a simulated stellar spectrum with 
a `noisy' wavelength solution with an error of 0.0008~{\AA} in each pixel to a perfect stellar 
template with the `correct' wavelength solution. The final RVs are shown as black filled circles, 
with an rms value of 0.4\ms. If we use the `correct' wavelength solution for all the simulated spectra, 
the final RVs are zeros, which are shown as red open circles here. No noises are added to 
the simulated stellar spectra.  
\label{fig:rv_error_wavelengthsolution}  } 
\end{figure}

As described in section 2.3, our derived wavelength solution has a typical wavelength solution error of $\sim$0.0008~{\AA} 
for each pixel. 
This wavelength solution error would lead to RV measurement uncertainties. We conducted simulations to study its resulting RV error.
First, we simulate stellar spectra using an input barycentric velocity variation from -30 to +30~km~s$^{-1}$ 
in steps of 5~km~s$^{-1}$ with a spectral template of Tau Ceti and a `correct' wavelength solution. 
We then create `noisy' wavelength solutions by injecting Gaussian distributed errors 
with $\sigma\sim$0.0008~{\AA} onto each pixel of the `correct' wavelength solution. 
By matching the simulated spectra having `noisy' wavelength solutions to the perfect stellar template with a `correct' wavelength solution, 
the RV error induced by the `noisy' wavelength solution can be derived. The results of the simulation are shown in 
Figure~\ref{fig:rv_error_wavelengthsolution}. We do not add any Poisson photon noises to the 
simulated stellar spectra to ensure the derived RVs only contain effects from the wavelength calibration error, which have an rms of $\sim$0.4\ms. 
This rms value of $\sim$0.4\ms represents the RV error induced by the wavelength calibration error, 
which is included in Table~\ref{tbl:budget}. 

\subsection{Micro-Telluric Line Contamination}
Another source of RV error is the Earth's atmospheric contamination, which creates telluric line features on 
stellar spectra obtained on the ground. Although telluric lines are known to be strong in the infrared bands 
(beyond $6000$~\AA), there exists many micro-telluric lines in the optical band 
\citep[4500$-$6000~\AA;][]{cunha14, smette15}. Since these stellar absorption lines move against 
micro-telluric lines when the star moves relative to the observatory, these lines 
will produce systematic noises when measuring stellar line movements. We masked out parts of the stellar 
spectra where telluric lines are deeper than $1\%$ during the RV extraction step. We also make the mask bigger to 
take into account the fact that these lines can move by as much as $\pm$30~\kms per year due to the barycentric motion, 
which can make some spectral orders useless when deriving RVs.
However, according to the simulation study by Sithajan et al. (2018, in preparation), micro-telluric lines 
shallower than $1\%$ in the optical band can still create an RV error around 0.1$-$0.2\ms, which is included in 
Table~\ref{tbl:budget} as an error from the micro-telluric line contamination. 
The size of this error is mainly dependent on stellar spectral type and the Earth's atmospheric 
water vapor column density distribution during the observation \citep{cunha14}. 




\subsection{Instrument Drift Correction Error}


We estimated the instrument drift correction error based on the combined  error of the instrument long-term drift calibration error and the intra-night drift calibration error. For the long-term drift calibration error, we estimate that the error of each instrument drift measurement using one single ThAr exposure from the `reference lamp' is on the order of $\sim$0.3\ms, 
which mainly comes from the photon counting errors of the Thorium emission lines used in the calculation. 
We took three `reference lamp' exposures every night, and found that the drift measurement from the first exposure is usually off 
by 1 to 2\ms comparing with the next two measurements. Thus, we decided to skip the first exposure, and use 
the second and third exposures from the `reference lamp' to calculate the nightly instrument drift, which 
results in an instrument long-term drift correction error of $\sim$0.2\ms. We currently do not know why 
the first exposure shows poor performance for RV drift measurements. We have verified that thorium line intensities did 
not change significantly from the first to the third exposure. We thus can only speculate that it likely 
occurred because the lamp was only turned on an hour ago and still warmed up.

\begin{figure}
\includegraphics[angle=0,width=8.2cm]{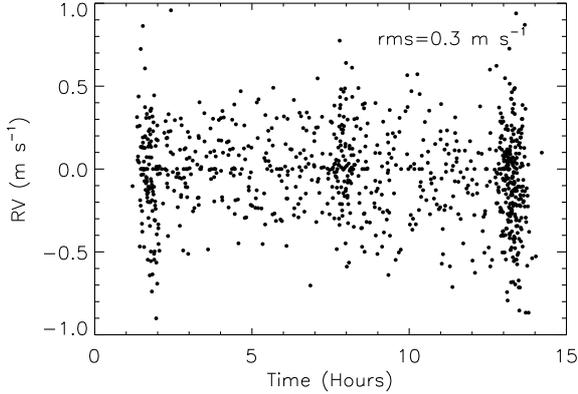}
\caption{Intra-night instrument drift residuals tracked using the ThAr `operation lamp'.  We first fit a linear function to these measurements from 
the `operation lamp' on each night, then remove this best fit to calculate the residuals. The rms of the residuals is $\sim$0.3\ms. 
This best linear fit from each night will be used to derive the instrument drift corrections for stellar exposures on that night.  
\label{fig:drift}
} 
\end{figure}

For the intra-night drift correction error, we estimate the error from the residual after we took the linear fit to each night 
instrument drifts as illustrated in Figure~\ref{fig:drift2} and subtracted the linear trend in instrument drifts. Figure~\ref{fig:drift} 
shows the overall 
instrument RV drift residuals after removing the best linear fit from the instrument drifts measured 
using the `operation lamp' exposures on each of the 60~nights, with an rms 
value of $\sim$0.3\ms. This $\sim$0.3\ms is the error we adopt for the intra-night instrument drift correction. 
We can reduce this error down to 0.1\ms by committing more telescope time on each night to take 
more ThAr exposures. But we choose not to do that because the RV precision improvement is not 
significant, and more ThAr exposures on each night would yield less stellar exposures.

By combining the error of $\sim$0.2\ms for nightly instrument drift correction and $\sim$0.3\ms for intra-night instrument drift correction, 
we have the total instrument drift correction error on the order of $\sim$0.36\ms for each stellar exposure, which is included 
in Table~\ref{tbl:budget}. 

\begin{figure}
\includegraphics[angle=0,width=8.2cm]{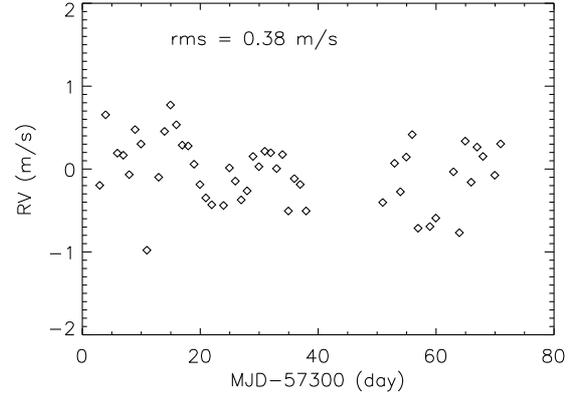}
\caption{Investigation of the instrument drift correction error. We use both ThAr lamps to track the instrument drifts 
over 60~days. The residuals after one drift is subtracted by the other drift is shown in this figure. 
The rms of the residuals is $\sim$0.38\ms.  \label{fig:drift_subtract}
} 
\end{figure}

There is a second method to estimate the instrument drift correction error. We conducted an experiment in which we used 
both of the `reference lamp' and `operation lamp' to track the same long-term instrument drifts. During the experiment, we took 
sequential exposures from both lamps separated by less than 10mins on each night, so the nightly instrument drifts between these 
exposures are negligible. We  subtracted one instrument drift from the other, and the result is shown in Figure.~\ref{fig:drift_subtract}. 
The rms of the residuals is 0.38\ms, which is very close to the theoretically estimated value of 0.36\ms introduced above. 
This experiment demonstrates the use of 0.36\ms as our instrument drift correction error in Table~\ref{tbl:budget} is reasonable. 


\subsection{Barycentric velocity correction}


We estimated barycentric velocity correction errors using simulations of stellar observations incurred 
from errors in the stellar coordinate ($\sim$3~mas), observatory coordinate and altitude ($\sim$10~meters), 
proper motion ($\sim$1~mas/yr), parallax ($\sim$0.2~mas), and timing error of the flux weighted exposure centroid time ($\sim$1~s). 
The accumulated error is $<$0.05\ms for Tau Ceti, which is also included in Table~\ref{tbl:budget}. 

However, for a fainter star with a longer exposure time, this barycentric velocity correction error is larger, and
 is dominated by the timing error of the flux weighted exposure centroid time. 
We use simulations to estimate this timing error for fainter stars. For 
a star with $V=9$ magnitude, the stellar PMT count rate is $\sim$10~counts/s, including the background noise level 
of 2.4 counts/s for the PMT. Under variable weather conditions, 
our simulations show the timing error can reach  $\sim$11~seconds for a 50~min exposure of a $V=9$ star. 
This would generate a barycentric velocity correction error as large as $\sim0.2$\ms.

\subsection{Template Interpolation Error}
We use a spline function to interpolate the stellar template spectrum at certain wavelengths, 
which is the best guess using all the information from the observation. This can introduce 
mathematical errors. We estimate these errors by carrying out a simulation using just 
one absorption line. A Gaussian function is used to represent this absorption 
line. We then create 121 spectral lines to represent the same spectral line with 
barycentric velocity movements from -60 to +60~$\rm{km\;s^{-1}}$. 
For this Gaussian-shaped narrow spectral line, the interpolation brings an error on the 
order of 2\ms for TOU's spectral resolution and pixel sampling of each resolution element. 
If the absorption line profile is broader due to fast rotation of a star, this interpolation 
error is smaller because the line is better sampled in the pixel space. Thus, the sharper 
the line is, the larger this interpolation error is for each line. 
Since this error is only dependent on the pixel position of the line center, it will average out 
with hundreds of absorption lines in each spectral order. Since we use more than $\sim$400 stellar absorption 
lines to calculate RVs, this error is likely smaller than 0.1\ms, which is included 
in Table~\ref{tbl:budget} as the template interpolation error. 

\subsection{PSF Variation \label{s:psf}}
\begin{figure*}
\includegraphics*[angle=0,width=8.2cm]{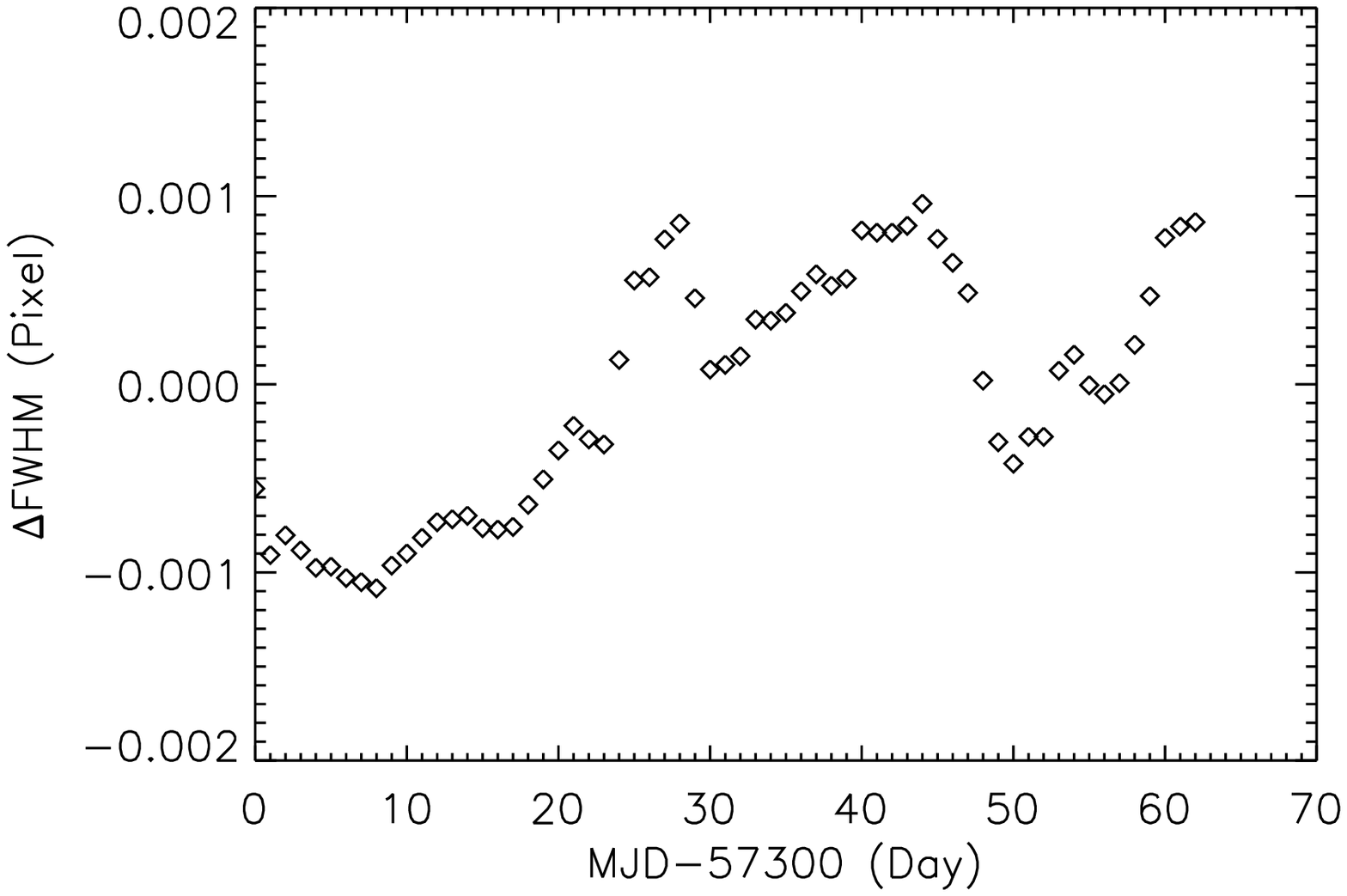}
\hfil
\includegraphics*[angle=0,width=8.2cm]{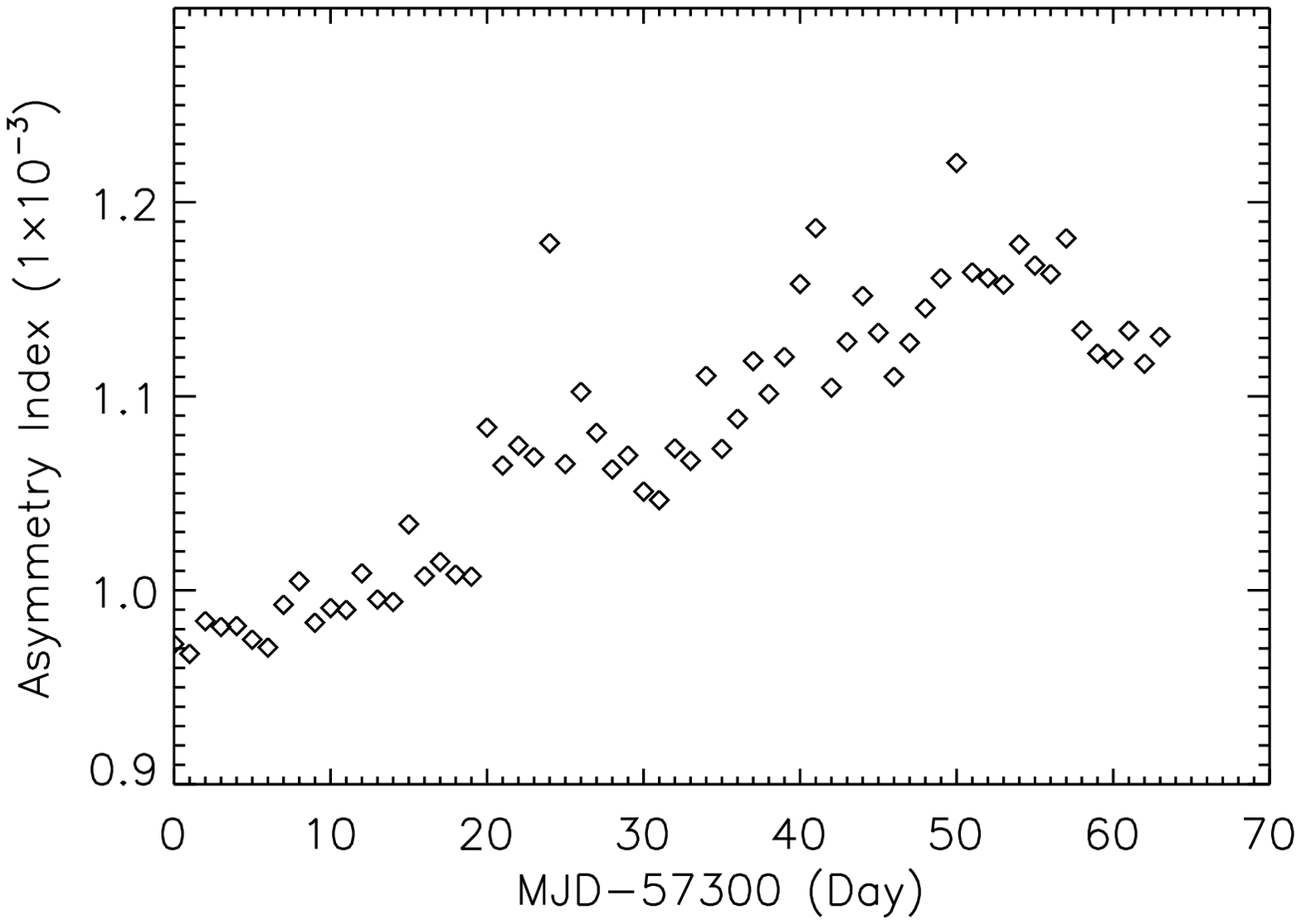}
\caption{Illustration of PSF shape evolution with time. The left panel displays variations of 
FWHM  (in pixels)  of the PSF with time. The right panel displays variations of the asymmetry index 
as defined in Section~\ref{s:psf} with time. The FWHM is stable within 0.001~pixel, and the asymmetry index is 
stable within $1\times10^{-4}$~dex.
\label{fig:psf_change1}
} 
\end{figure*}

In our pipeline, we choose not to use deconvolution to correct variations in our instrument PSF, which 
are usually induced by unforeseeable optics or optical illumination changes due to instrument temperature 
and pressure variations and/or fiber coupling. Although the environment of the TOU instrument is very well controlled, 
there still exist minor variations, which affect RV extraction in the spectrum matching process 
and produce RV measurement uncertainties. We use a set of 400 strong and un-saturated Thorium emission lines to trace 
the instrument PSF variations. We use a function that contains three Gaussians to fit each Thorium line. This three-Gaussian function 
profile contains a large central Gaussian function and two small Gaussian functions with their centroids $\pm1$ pixel away from 
the central Gaussian function. The full width at half maximum (FWHM) is derived using the central Gaussian function. 
We also defined an asymmetry index, which is the area difference between the left and the 
right Gaussian functions divided by the total area under all three Gaussians. 
Figure~\ref{fig:psf_change1} shows the variations of PSF FWHM and asymmetry index during two months of observations. 
These plots show that the FWHM is stable within 0.001~pixel, and the asymmetry index is stable within $1\times10^{-4}$~dex.

The PSF variation with time can introduce systematic errors in RV measurements by 
changing the shapes of stellar lines. These spectral line profile variations can be interpreted as 
RV shifts when matching with a stellar template. Using the measured PSF variations, 
we have simulated their effects on TOU RV measurements for Tau Ceti. 
Our results show that PSF variations over time can cause an rms of $\sim$0.2\ms RV error, which is included 
in Table~\ref{tbl:budget} as an error caused by PSF Temporal Variation.
 

\begin{figure}
\includegraphics[angle=0,width=8.2cm]{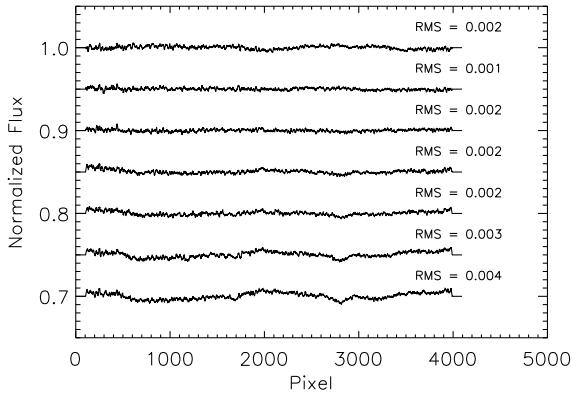}
\caption{ Evolution of normalized flat-fielding curve over time for the TOU spectrograph order 110. From 
top to bottom, there is an increment of 10 days between each flat-fielding curve. An offset of 0.05 unit of 
normalized flux is added between each curve for display purposes. The flat-fielding curves are derived using 
high SNR calibration spectra of Tungsten lamp on each observation night. The rms value of each normalized 
curve is also shown in the figure. \label{fig:blaze}
} 
\end{figure}

\begin{figure}
\includegraphics[angle=0,width=8.2cm]{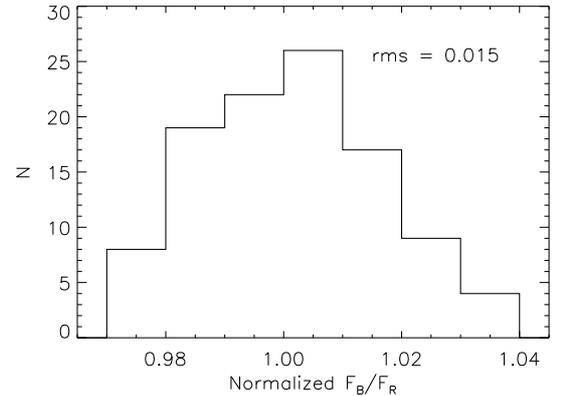}
\caption{ Histogram distribution of the normalized flux ratios between the blue orders and the red orders from the TOU exposures of 
Tau Ceti. The normalized flux ratios are calculated between the total photons from order 105$-$110 and order 125$-$130, then normalized 
by the median value of all the flux ratios. The normalized flux ratios have an rms value of 1.5\%. 
\label{fig:color}  } 
\end{figure}



\subsection{Flat-Fielding Variation}
In this section, we will study the impact of flat-fielding variations on RV measurements. Since the flat-fielding modulates with 
the flux intensity of a stellar spectrum in each spectral order. 
%
For a perfectly stable instrument, the effective flat-fielding function for each echelle order should be constant 
over time. However, it is never constant due to some instrument effects, which can cause a problem when 
we match observed stellar spectra with a spectral template. Figure~\ref{fig:blaze} shows the flat-fielding 
variations during our two months of Tau Ceti Observations. The relative value (rms) changes 
by $\sim$0.1$-$0.4$\%$. The best method to correct this small flat-fielding variation is to take 
sufficient calibration data on each observation night and generate 
high SNR ($\sim$1000 per pixel) flat-fielding spectra for correction. 

Our simulations show that flat-fielding variations of $\sim$0.4\% can induce up to 0.4\ms 
RV errors for G-type stars, and 0.2\ms RV errors for K- and M-type stars in the optical band.
The  impacts are bigger for G-type stars than K- and M-type stars because spectra of K- and M-type 
stars contain more RV information than those of G-type stars and the flat-fielding variation effect 
can be averaged out by some degree when assuming all spectra have the same SNRs \citep{bouchy01, wang11}.
High SNR calibration data from the Tungsten lamp allow us to correct most of the flat-fielding variations, 
which leaves $<0.1$\ms RV noise. This error is included in Table~\ref{tbl:budget} as an RV error from the 
flat-fielding variation.
 


\subsection{Color Effect}

Because of the atmospheric dispersion and extinction effect at different airmass between 
many exposures of the same star, we anticipate the flux weighted centroid time of observation, 
which is used to calculate the barycentric velocity correction, is color dependent. 
Since we optimized the schedule to observe Tau Ceti at a low and similar airmass each night, 
the atmospheric dispersion effect is minimal. 
However, the light transmission through the atmosphere is also strongly dependent on wavelength and 
airmass due to the atmospheric extinction. 
Therefore, the flux weighted centroid time of a stellar exposure, and the barycentric velocity correction 
calculated using the flux weighted centroid time, is also wavelength dependent. This can create an RV 
error smaller than 0.1\ms for a 30~min exposure using 3800$-$6800~{\AA} according to the simulation work of 
\citet{blackman17}, in which the authors suggest using a multiple-channel exposure meter to correct this effect. 

Here we estimate the impact of this effect on our RV measurements of Tau Ceti from TOU. 
The flux weighted centroid time in the blue spectral orders ($t_B$) and red spectral orders ($t_R$) 
are slightly different due to the atmospheric dispersion/extinction. When calculating errors 
of relative RV measurements, we only need $\Delta(t_B-t_R)$, which can be estimated using the 
flux distribution variation from the red to blue spectral orders of the Tau Ceti exposures. 
We first calculate the flux ratios between the red spectral orders and blue spectra orders $F_B/F_R$ using 
the total photon numbers from order 125$-$130 ($F_B$) and order 105$-$110 ($F_R$), then 
normalize them using the median value of all the $F_B/F_R$.
The normalized ratios, hereafter $\overline{F_B/F_R}$,  are shown in Figure~\ref{fig:color}, from which 
$\Delta(\overline{F_B/F_R})$ is derived and has an rms value 1.5\%. 
The amplitude of this variation mainly depends on the airmass of the target during the observation, the extinction law, 
and the total exposure time \citep{blackman17}. 
The flux weighted centroid time difference between the blue order and red order $\Delta(t_B-t_R)$ can be expressed as 
$\rm \sim 0.21\times \Delta(\overline{F_B/F_R})\times Exposure\;Time$. The factor of 0.21 comes from the fact that in an extreme case where the 
transmittance in the blue order changes linearly with time from $\rm T_{Initial}$ to 0 due to atmospheric dispersion change
while the transmittance in the red order does not decrease at all, thus $\Delta(\overline{F_B/F_R})=1$, this flux weighted centroid time 
would decrease by a 21\% of the total exposure time from the red order to blue order. 
With the rms value of $\Delta(\overline{F_B/F_R})$ as 1.5\%, this results in a 1.9~seconds rms of centroid time difference from a 10~min exposure of Tau Ceti, 
which can induce an error of 0.035\ms (rms) when measuring relative RVs. 
Our result estimated from real observation data is similar to that from the simulation work of \citet{blackman17}, 
in which they would predict an RV error smaller than $\rm 0.15\times \frac{10~mins}{30~mins}\sim0.05$\ms for a 10 min exposure. 
Since we use three consecutive 10 min exposures of Tau Ceti to calculate one RV point, the error from this color effect is $0.035\times3^{0.5}=0.06$\ms, 
which is included in Table~\ref{tbl:budget} as an error from the color effect.

\subsection{CCD Effect }
If a spectral line passes over an imperfection of the CCD, this spectral line can undergo a deformation to
introduce an RV variation that is correlated or anti-correlated with the Barycentric velocity. This effect can potentially 
create a false one year periodical RV signal \citep{dumusque15b}. We have not been able to identify 
such a one year period signal in our TOU data, thus we are not able to estimate the amount of error due to the CCD 
imperfections in this study.

\citet{blake17} simulated the impact of inefficient charge transfer of CCD, called charge transfer inefficiency (CTI), 
on RV measurement precision. If CTI is constant over time for both stellar spectra and wavelength 
calibration spectra, this effect is not important. However, since CTI changes as a function of flux level, it can introduce 
a source of systematic RV measurement error. This systematic error can be as large as 0.3\ms for a 
$\rm \Delta CTI = 10^{-7} \; pixel^{-1}$ \citep{blake17}. The TOU CCD is a Fairchild 4k$\times$4k back-illuminated 
CCD detector which has a nominal CTI$\rm <5\times10^{-6}\; pixel^{-1}$. 
Based on the analysis of  \citet{bouchy09}, our best guess of the RV measurement error for bright stars 
from CTI for TOU is smaller than 0.1\ms, which is included in Table~\ref{tbl:budget}.

\subsection{Telescope Guiding}
Another error source is the telescope guiding. 
The small image movements feeding into the fiber tip can cause tiny spectral line shifts on the CCD image, 
which translates to RV noises. Our telescope has a measured guiding error of $\sim$0.3~arcsec (rms value).
With a mode scrambling gain factor of 2000 from the science fiber with an optical mode double scrambler \citep{ge14a} 
used in the TOU spectrograph, this guiding error can lead to stellar absorption line shifts on the scale of 3$\times10^{-4}$ 
pixel on the CCD. These randomly generated line shifts from telescope guiding error can produce an RV error on the 
order of $0.3$\ms for TOU, which is summarized in Table~\ref{tbl:budget} as the RV error from telescope guiding. 


\subsection{Stellar Activity Jitter}


Stellar activity, including stellar oscillations, granulation, spots, and plages, can affect RV 
measurements \citep{dumusque15a}. However, it is hard to directly measure this error as it is usually 
mixed with other measurement uncertainties. \citet{wright05} studied stellar RV jitters from RV 
measurements of $\sim$450 FGK stars from Keck Observatory and concluded that RV 
jitter amplitude depends on stellar activity levels and spectral types, varying from $\sim$1\ms to a few tens \ms.
\citet{isaacson10} studied chromospheric activity for more 
than 2600 main-sequence and subgiant stars in the California Planet Search (CPS) program and showed that solar-type stars 
typically have a few\ms RV noise induced by stellar activity. 
\citet{mascareo17} also studied the RV jitter induced by stellar activity using HAPRS data for 55 FGKM dwarf stars and 
were able to provide jitter measurements and fitting down to the sub\ms level. 
Using the equation provided by \citet{mascareo17} for G-type stars, we derive an RV jitter of $\sim$0.45\ms for Tau Ceti 
using $\rm \log_{10}(R_{HK}^{'})=-4.98$ from \citet{isaacson10}, which is included in Table~\ref{tbl:budget} as stellar jitter. 

\section{Summary and Discussion}
In this paper, we present our data reduction pipeline for the TOU high-resolution spectrograph and its RV performance. 
To understand the RV error budget when observing bright targets with the TOU spectrograph, 
we also conducted a detailed study using a bright RV stable star, Tau Ceti, as an example. 
Our results indicate that TOU has reached a \sms RV precision, which can be used to detect close-in super-earths around 
nearby bright stars and monitor their stellar activity properties. The TOU spectrograph was relocated to the UF 50-inch 
robotic telescope, called the Dharma Endowment Foundation Telescope (DEFT), at Mt. Lemmon in 2015. 
From 2016-2021, TOU is mainly used for the high cadence and high precision Dharma Planet Survey \citep[DPS,][]{ge16, ma18a} 
and the TESS \citep[Transiting Exoplanet Survey Satellite,][]{ricker15} planet candidates follow-up. 

As illustrated in our error budget discussions in Section~\ref{sec:error}, major efforts are needed 
to achieve 0.1\ms precision with the next generation Doppler RV instruments. This includes, but are not limited to, 
 improving the PSF deconvolution and micro-telluric line removal, deriving a better wavelength calibration solution, building a 
more stable instrument, and getting a better understanding of stellar activity jitter and its removal. 

Our study shows that wavelength dispersion of a high resolution optical spectrograph 
must be known with an incredible precision to deliver the ultra-high RV precision necessary for the detection of an Earth analog.
Due to the limitations of the currently widely used emission lamps (such as ThAr lamps) and absorption cells 
(such as iodine cells) as wavelength calibration sources \citep{butler96, fischer16}, many research groups are 
actively developing new calibration sources to reach better than 0.1\ms calibration precision in the optical band. 
These calibration sources include laser-frequency combs \citep{li08, steinmetz08}, Fabry-Perot interferometer 
calibration sources \citep{wildi11,schwab15, bauer15}, and the SINE source developed by our group \citep{wan10, ge14b}. 
These new calibration sources pose significant improvements of the wavelength calibration from the currently 
widely used ThAr lamps and iodine cells with precise RV instruments, 
and make it possible to reduce the wavelength calibration error to the 0.1\ms level or better. 
These new calibration sources can also improve the instrument drift correction efficiency 
by reducing the calibration exposure time needed on each night. This is because there are more strong and stable 
lines from these new calibration sources available to calibrate instrument drifts, comparing with only several 
hundred strong Thoriums lines available to use from a ThAr lamp. 

The stellar activity jitter, including but not limited to, RV perturbations induced by stellar spots, plages, and granules, 
is probably the strongest limitation for Doppler measurements aiming at the \sms precision. 
It would be even more challenging to detect an Earth-twin in the habitable zone around a sun-like 
star as the RV signal caused by the planet is on the order of $\sim$0.1\ms which is much smaller than 
typical stellar noises \citep[e.g.][]{wright05, isaacson10, mascareo17}. 
In order to correct RV jitters caused by stellar activities, it is important to study stellar activities and mitigation techniques. 
On the other hand, our Tau Ceti measurements 
show that this star has a relatively low stellar jitter level of $\sim$0.45\ms. This suggests that it is 
quite possible that some solar type stars may be very quiet in stellar activities, 
which can be ideal targets for searching for Earth analogs in the future.  
Since stellar activities from a star have a certain temporal structure, one way to tackle 
this stellar activity noise problem is to acquire as many RV data points as possible to mitigate the noise. 
This process can be done simply by using an advanced observation scheduling scheme like the one proposed by 
\citet{dumusque11}. Another way to reduce the stellar jitter impact is to use a better stellar jitter model, 
like time-dependent moving average models \citep{tuomi13} or wavelength-dependent noise models \citep{feng17}, 
both of which require larger than usual RV datasets to constrain the jitter model. Imaging or photometry 
data can also help remove some of the stellar jitter noise. Since activity jitter can be color dependent, 
\citet{ma12} proposed a multi-band RV technique to remove activity jitter induced by stellar spots and 
detect planet signals. \citet{dumusque15a} studied the Sun using HARPS-N 
and developed a technique to extract its true RVs using full-disk photometry data. \citet{dai17} used a 
Gaussian process regression based on both RV and photometric data to successfully detect a planet signal 
in addition to stellar activity jitter noises. \citet{davis17} proposed a statistical technique to disentangle stellar 
activity jitter from planet signals using principal component analysis (PCA). 

For the errors caused by micro-telluric lines, we cannot use the 
mask method \citep{pepe02} to reduce them because there are too many micro-telluric lines to mask. There are three possible 
methods to reduce these errors. 
The first one is to launch the spectrograph to space together with an optical/IR telescope, which is expensive. 
It is also difficult to maintain the stability of the spectrograph in space. The second way is to create synthetic 
models of micro-telluric lines using weather data collected during observations \citep{seifahrt08, bean10, kausch15}, 
then apply these synthetic models to remove telluric lines from stellar spectra. This method strongly depends on the completeness of 
the telluric line list, the accuracies of the telluric line parameters, and the accuracies of air molecular column densities 
(such as $\rm H_2O$ and $\rm O_2$) measured during observations \citep{gullikson14, bertaux14}.  
The last method is to collect spectra from early-type (B- or A-type) telluric standard stars to derive empirical models of
micro-telluric lines \citep{vacca03}. 
To obtain an accurate telluric line model, observations of the telluric standard star and the science target
should be conducted around the same time and airmass to ensure similar light paths through the atmosphere 
for both stars. Such scheduling is not always easy to arrange, and a significant amount of telescope time is 
needed to acquire high SNR ($>1000$ per pixel) spectra from telluric standard stars. 

A better optimized schedule of observations can also improve the performance of next generation of RV instruments. 
Achieving the same SNR for each exposure of the same star can help reduce the error from CTI \citep{blake17}, because 
CTI is known to be a function of the flux level. 
Nevertheless, the best way to minimize this CTI effect is to arrange the CCD orientation to let its readout direction perpendicular to the 
dispersion direction like what we did for TOU.
Repeatedly observing the same star under the same airmass condition 
can reduce the RV noises from color effect. These errors are negligible for now when targeting $\sim$1\ms RV precision, 
but will be significant for the next generation of instruments which are targeting $\sim$0.1\ms RV precision.  

Major efforts have been made by us to develop and test the data reduction pipeline for the TOU spectrograph, 
which has delivered a \sms RV precision. Our pipeline study using the TOU spectrograph demonstrates it 
is necessary to carefully go through all the steps in the data reduction to fully understand all error sources which 
contribute to RV performance on stars. Only after we understand all of the RV error sources and can 
make precise correction, there is a possibility of achieving 0.1\ms using a stable and very high-resolution optical spectrograph.

\section*{Acknowledgments}
We would like to thank Dr.Gill Nave at NIST for the help of calibrating our Thorium-Argon lamps 
using their ultra-high resolution Fourier Transform Spectrograph, 
Frank Varosi from the University of Florida for his kind help on studying the instrument stability, and Nolan Grieves 
from the University of Florida for his kind suggestions to improve this paper. 
We thank the referee for the constructive suggestions which have helped improve the paper quality.
We would also like to thank Matthew Muterspaugh and Michael Williamson from Tennessee State University for their help TOU 
observations using the 2-m Automatic Spectroscopic Telescope at Fairborn Observatory.
The Dharma Planet Survey was supported by the Dharma Endowment Foundation. 
Bo Ma thanks the support of the NASA-WIYN observation award. This research has made 
use of the Exoplanet Orbit Database and the Exoplanet Data Explorer at exoplanets.org.

\bsp	
\label{lastpage}
\end{document}